# Unified Growth Theory Contradicted by the Economic Growth in Africa

Ron W Nielsen[1]

Environmental Futures Research Institute, Gold Coast Campus, Griffith University, Qld, 4222, Australia

December, 2015

One of the fundamental postulates of the Unified Growth Theory is the claimed existence of three distinctly different regimes of economic growth governed by three distinctly different mechanisms of growth. However, Galor also proposed that the timing of these regimes is different for developed countries and for less-developed countries. Africa is the perfect example of economic growth in less-developed countries. The data used by Galor, but never properly investigated, are now analysed. They turn out to be in dramatic contradiction of this theory.

**Introduction**

There is no science without data but there is also no science without scientific analysis of data. Many attractive theories and explanations can be formulated but if they are not based firmly on rigorous analysis of data they are only, at best, just interesting stories. They may contain elements of truth but folklores of many cultures are full of such stories and they also contain elements of truth. Fantasy and leaps of faith might be inspiring and productive even in the scientific research but they have to be soon tested by the scientific process of investigation.

However, if one leap of faith is followed by another, if one fantasy creates another, then we no longer deal with science but with fiction. It is then easy to loose scientific perspective and defend emotionally the widely-accepted dogmas, based on faith.

Any theory that cannot be checked by data is unscientific even if it is based on scientifically attractive ideas. Such a theory has to be put aside until it can be checked by relevant data. Even if a theory is confirmed by many sets of data it can be still challenged by a single set of contradicting data. Any theory contradicted by just one set of good data has to be either revised or rejected. Any research, any intellectual activity, which ignores these fundamental principles of scientific investigation is unscientific even if it is intellectually stimulating and attractive.

In science it is important to look for data confirming theoretical explanations but it is even more important to discover contradicting evidence, because data confirming a theory confirm only what we already know but contradicting evidence may lead to new discoveries.

Thus, for instance, if scientific analysis of data is found to be in agreement with the Unified Growth Theory, this theory may then be considered as being confirmed by data and its

---

explanations of the mechanism of the economic growth may be then accepted. However, if just one set of data is found to be in contradiction with this theory, then this theory can no longer be accepted in its present form. It has to be then either modified to bring it in agreement with data, or rejected if such modification is impossible. There is no scientific gain in accepting such a theory. On the contrary, its continuing acceptance is detrimental to science.

When the incorrect theory is rejected we can then look for a better explanation of the mechanism of the economic growth, explanation, which might open a new line of research and which might help to understand not only the past but also the present economic growth and maybe even its future. There are no sentimental values in scientific research and no emotional attachments, and any scientist should be prepared to have his or her explanations challenged by science.

Galor's concept of the three regimes of growth, the concept he inherited from his many predecessors, is both harmful and misleading. It creates a sense of security when there is none.

According to this concept, after a long stage of economic stagnation, which lasted for many thousands of years, we have now managed to escape from the mythical Malthusian trap and we can, at last, enjoy a sustained economic growth. The opposite is true. The economic growth in the past was steady and secure (Nielsen, 2015a) but is has now reached an insecure stage (Nielsen, 2015b, 2015c), which requires close monitoring and control. The Unified Growth Theory is not only unscientific (Nielsen, 2014) but also potentially harmful.

**Galor's three regimes of growth**

One of the fundamental postulates of the Unified Growth Theory (Galor, 2005, 2011) is that the historical economic growth in various countries and regions can be divided into *three distinctly different regimes of growth* governed by distinctly different mechanisms of growth.

These alleged regimes are:

1. The regime of Malthusian stagnation. This regime lasted allegedly for thousands of years and was characterised by random fluctuations and oscillations around a stable Malthusian equilibrium. Galor claims that this epoch of stagnation commenced in 100,000 BC (Galor 2008, 2012). Scientific justification for this claim is unclear because Maddison's data (Maddison, 2001), which Galor used (but never analysed) during the formulation of his theory extend only down to AD 1. Extending the alleged epoch of stagnation to 100,000 BC sounds like a large leap of faith. However, the same data, when analysed, demonstrate clearly and convincingly that the three regimes of growth did not exist (Nielsen, 2014) at least for the world economic growth and for the growth in Western Europe. Here we have the theory contradicted not just by one but by two sets of data, which in science is sufficient for the postulate to be rejected.

    Galor claims that the regime of Malthusian stagnation was terminated in 1750, or around the time of the Industrial Revolution, 1760-1840 (Floud & McCloskey, 1994), in developed countries and in 1900 in less-developed countries (Galor, 2008, 2012). How he managed to determine these dates is unclear. The three regimes of growth did not exist for the world economic growth and for the growth in Western Europe (Nielsen, 2014) and we shall soon demonstrate that Galor's claim is also contradicted by the data describing economic growth in less-developed countries.

2. The post-Malthusian regime. According to Galor (Galor, 2008, 2012), this mythical regime was between 1750 and 1870 (overlapping the time of the Industrial Revolution) for developed countries but it commenced much later, in 1900, for less-developed countries and it still continues.
3. The sustained-growth regime. According to Galor (Galor, 2008, 2012), this regime commenced in 1870 for developed countries and it still continues.

The claim of different timing for the postulated distinctly different regimes of growth is an integral part of the Unified Growth Theory and is expressed in two other fundamental postulates: the postulate of the differential takeoffs and the postulate of the great divergence, all supported by impressions and by the incorrect interpretation of data. When closely analysed, all these postulated are contradicted by data used by Galor.

In the discussion presented in this publication we shall focus on the less-developed countries and we shall demonstrate that Galor's claim of the two regimes of growth in these countries is dramatically contradicted by data, coming precisely from the same source as used during the formulation of his theory. Even though Galor had access to these excellent set of data he has not analysed them but was obviously guided only by impressions, which is most unfortunate because the mathematics of the historical economic growth is so elementary and so simple (Nielsen, 2014, 2015a).

**Historical economic growth**

The latest and updated compilation of data describing historical economic growth was published by Maddison in 2010 (Maddison, 2010). These are virtually the same data as published earlier (Maddison, 2001) and used (but not analysed) by Galor (2005, 2011) but in the 2010 edition the data were extended to 2008.

Recent extensive mathematical analysis of economic growth (Nielsen, 2015a) demonstrated that the historical economic growth, global and regional, can be well described using hyperbolic distributions. These distributions appear to be creating a significant problem with their interpretation. They are routinely seen as being made of two distinctly different components, slow and fast, jointed by a transition stage. However, their analysis becomes trivial if they are represented by their reciprocal values (Nielsen, 2014) because in this representation the confusing features disappear and hyperbolic distributions are represented by straight lines. We shall now use this method to analyse the economic growth in Africa.

**Africa**

Africa is a perfect example of a cluster of countries, which belong to the group of less-developed and least-developed countries. Out of the total of 48 least-developed countries in the world, 34 are in Africa (Bangla News, 2015; UNCTAD, 2013). With just one minor exception, Africa is made entirely of less-developed and least-developed countries (BBC, 2014; Pereira, 2011). The exception is Western Sahara, a small country in transition made of around 586,000 people (UNDATA, 2015).

Maddison's data for Africa serve, therefore, as an excellent source of information to test Galor's hypothesis of the existence of the distinctly different regimes of economic growth in less-developed regions. We shall demonstrate that this hypothesis is dramatically and clearly contradicted by the data.

Reciprocal values of data describing economic-growth in Africa are presented in Figure 1. Economic growth was clearly hyperbolic between AD 1 and around 1820 because the reciprocal values follow a straight line. There was definitely no stagnation. The concept of the regime of Malthusian stagnation is clearly contradicted by data. To prove its existence one would have to demonstrate a stagnant state of growth characterised by random Malthusian oscillation. The data contain no such evidence. On the contrary they show a convincing evidence of a steadily-increasing hyperbolic growth.

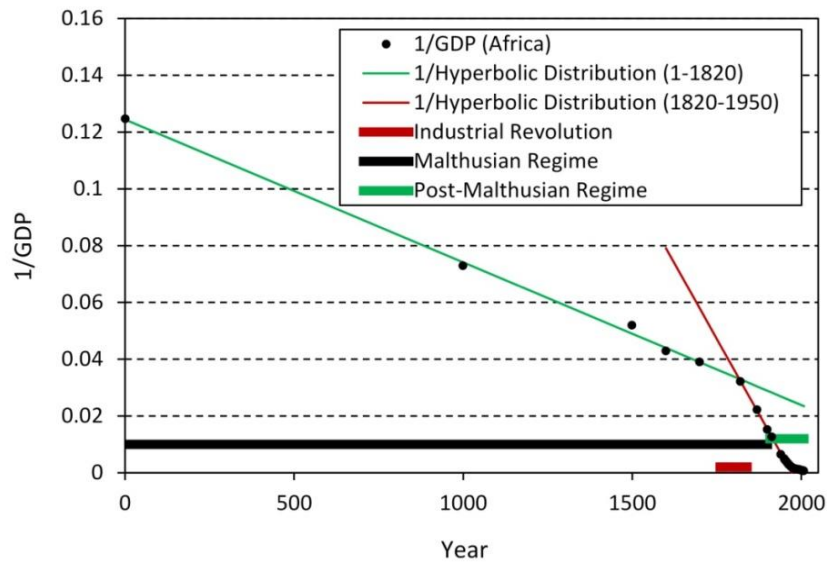

**Figure 1.** Reciprocal values of the GDP data (Maddison, 2010) for Africa compared with the hyperbolic distributions represented by the decreasing straight lines. The GDP is expressed in billions of 1990 International Geary-Khamis dollars. The two distinctly different regimes of growth postulated by Galor (2005, 2008, 2011, 2012) did not exist.

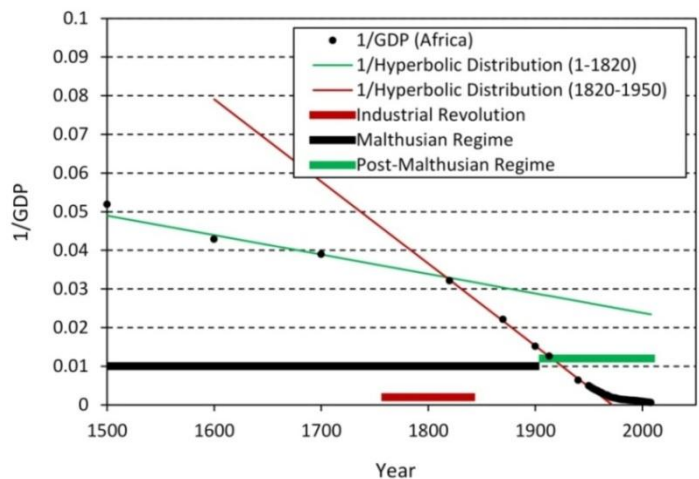

**Figure 2.** Reciprocal values of the GDP data (Maddison, 2010) for Africa between AD 1500 and 2008 compared with the hyperbolic distributions represented by the decreasing straight lines. The GDP is in billions of 1990 International Geary-Khamis dollars. The two distinctly different regimes of growth postulated by Galor (2005, 2008, 2011, 2012) did not exist. His postulate ignores data. There was no transition from stagnation to growth at any time, let alone at the end of the alleged Malthusian regime.

Furthermore, Galor's concept of Malthusian stagnation extending to 1900 ignores not only the data between AD 1 and 1820 but also the clear and dramatic transition, which occurred around 1820. It was *not* a transition from stagnation to growth but the *from growth to growth*, the transition from a slower but steady hyperbolic growth to a faster and steady hyperbolic growth. This pattern is in clear contradiction of the Unified Growth Theory (2008, 2012).

The concept of the regime of stagnation ignores not only the steady economic growth before 1820 and the dramatic change in the pattern of growth around that year but also the new hyperbolic pattern after 1820. The claim of Malthusian stagnation ending in 1900 for less-developed countries ignores also that absolutely nothing unusual had happened around that year. The economic growth continued undisturbed. The postulated Malthusian regime ends in the middle of nowhere. There is no justification for claiming the regime of Malthusian stagnation and no justification for terminating it in 1900 or at any other time.

The disagreement between the data and the story presented by Galor is already perfectly clear but there is still further contradicting evidence, which can be seen also in Figure 2. The commencement of the post-Malthusian regime in 1900 is totally unjustified because the data show absolutely no transition, no change of direction, around that time. In addition, the data demonstrate the existence of a feature, which is totally ignored by Galor: the diversion to a slower trajectory around 1950 indicated by the upward bending of the trajectory of the reciprocal values. So, while the data demonstrate a clear change in the pattern of growth, for Galor there is no change – just a continuation of his imaginary post-Malthusian regime. Data tell one story, Galor tells another, and in science data have the priority.

The disagreement between the postulates proposed by Galor and the data is also clearly demonstrated in Figures 3 and 4.

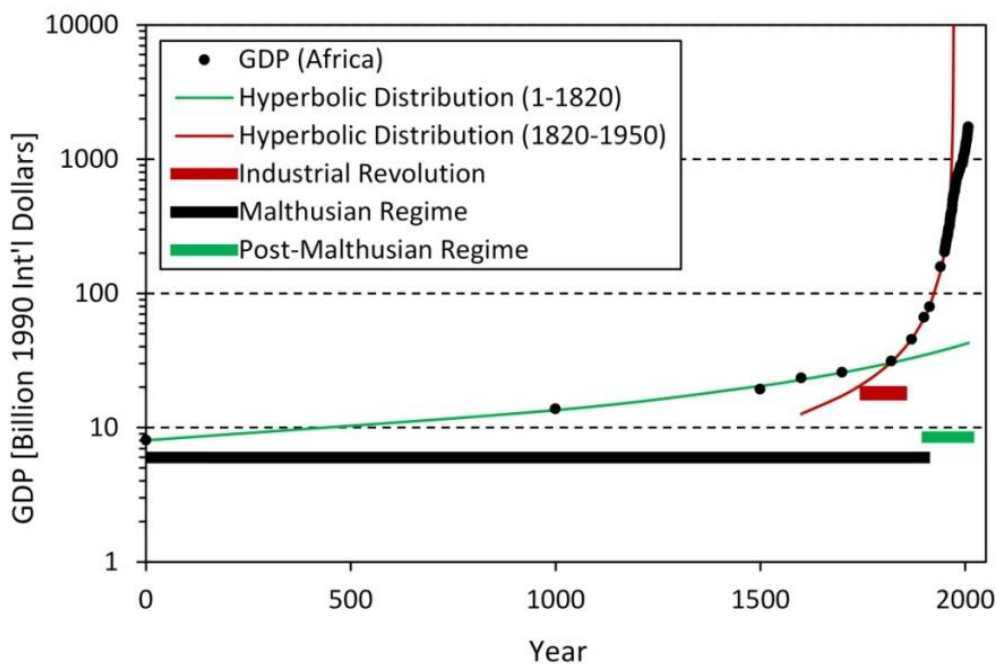

**Figure 3.** GDP data (Maddison, 2010) for Africa between AD 1 and 2008 compared with hyperbolic distributions. The GDP is in billions of 1990 International Geary-Khamis dollars. The two distinctly different regimes of growth postulated by Galor (2005, 2008, 2011, 2012) did not exist. His postulate ignores the data. There was no transition from stagnation to growth because there was no stagnation.

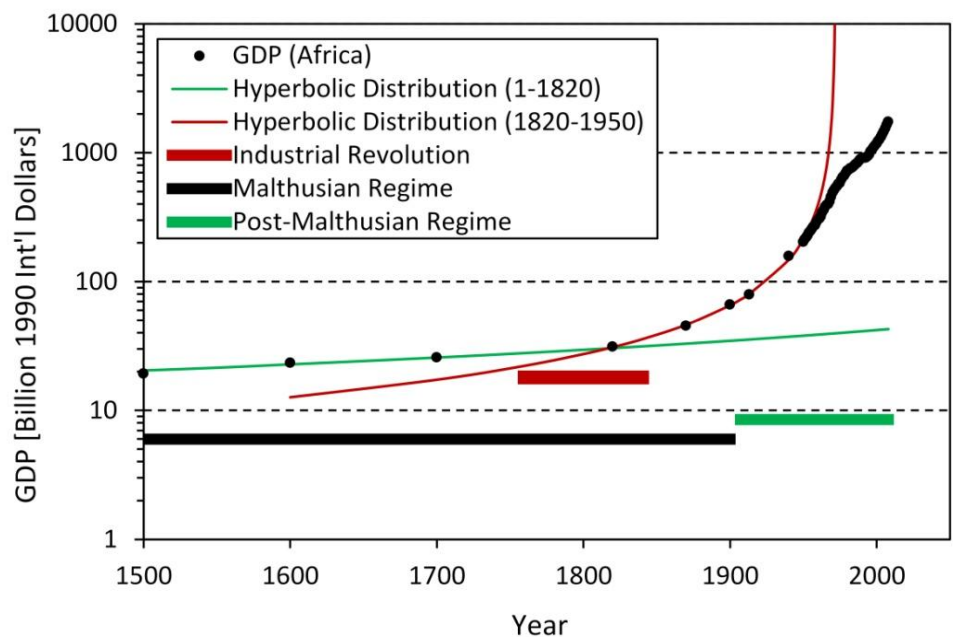

**Figure 4.** GDP data (Maddison, 2010) for Africa between AD 1500 and 2008 compared with the hyperbolic distributions. The GDP is in billions of 1990 International Geary-Khamis dollars. The two distinctly different regimes of growth postulated by Galor (2005, 2008, 2011, 2012) did not exist. His postulate ignores the data. The data are in clear contradiction of Galor's postulate. There was no transition from stagnation to growth because there was no stagnation.

Figure 3 shows that the Malthusian regime did not exist. The data show no evidence of a steady-state equilibrium, which is supposed to have characterised that mythical regime. In contrast, the data show a steady hyperbolic growth. Furthermore, the data presented in Figure 3 also demonstrate that this concept, as applied to less-developed countries (Galor, 2008, 2012), ignores the transition from slow to faster hyperbolic growth, which occurred around 1820. In addition, Figure 4 demonstrates clearly that the postulated epoch of Malthusian stagnation ends in the middle of nowhere. There is no change in the pattern of growth around the time of the alleged transition from the Malthusian regime to the post-Malthusian regime. However, there *was* a change during the imagined post-Malthusian regime but this change had been ignored by Galor. The economic growth was diverted from the hyperbolic growth to a slower trajectory.

**Summary and conclusions**

Galor proposed the existence of two regimes of growth for less-developed countries: the Malthusian regime, 100,000 BC – AD 1900 and the post-Malthusian regime, from AD 1900 (Galor, 2008, 2012). Economic growth in these countries is perfectly represented by the economic growth in Africa described by excellent set of data published by Angus Maddison (2001, 2010), the well-known and respected British economist. These data were known to Galor. He used their earlier version (but never analysed them) during the formulation of the Unified Growth Theory. The newer compilation (Maddison, 2010) must have been also known to him at the time of his proposed different timings of the mythical regimes of growth (Galor, 2012).

We have analysed the data for Africa and we have shown that they are in complete contradiction of Galor's postulate. Economic growth in Africa followed a hyperbolic distribution between AD 1 and 1820 and was converted to another, but faster, hyperbolic growth from 1820. From around 1950, economic growth was diverted from the hyperbolic growth to a slower but still increasing trajectory. In order to explain the mechanism of the economic growth in Africa we would have to explain why the growth was hyperbolic between AD 1 and 1820, why there was a transition to a new hyperbolic growth in around 1820, why the new economic growth was hyperbolic but faster after that year, and why the economic growth was diverted to a slower, but still-increasing trajectory. This is science *supported by data*.

The fiction, *contradicted by data*, is the story (Galor, 2005, 2008, 2011, 2012) revolving around his proposed different regimes of growth and around different mechanisms of growth during these mythical regimes. In particular, for the less-developed region discussed here and during the time of the availability of the mathematically-analysable data, the Malthusian regime did not exit and neither did the post-Malthusian regime. There was no stagnation and no transition to a new regime at the end of the postulated Malthusian regime. This alleged but non-existing regime ends in the middle of nowhere and ignores the transition from a slow to a faster hyperbolic growth, which occurred around 1820. The post-Malthusian regime commences in the middle of nowhere. It also ignores the transition to a slower trajectory, which commenced around 1950. There is absolutely no correlation between any of the two regimes proposed by Galor and the data. The data tell one story, the postulated regimes tell another story. Paradoxically, fundamental concepts of the Unified Growth Theory are contradicted by the data used, but never analysed, during its development. Any claim about the existence of Malthusian stagnation before AD 1 can be only based on unverifiable conjecture because there are no data describing economic growth during the BC era.

**References**

Bangla News, (2015). Bangladesh Chairmen of Least Developed Countries. http://www.en.banglanews24.com/fullnews/bn/115940.html

BBC, (2014). The North South Divide. http://www.bbc.co.uk/bitesize/standard/geography/international_issues/contrasts_development/revision/2/

Floud, D. & McCloskey, D.N. 1994. *The Economic History of Britain since 1700*. Cambridge: Cambridge University Press.

Galor, O. (2005). From stagnation to growth: Unified Growth Theory. In P. Aghion & S. Durlauf (Eds.), *Handbook of Economic Growth* (pp. 171-293). Amsterdam: Elsevier.

Galor, O. (2008). Comparative Economic Development: Insight from Unified Growth Theory. http://www.econ.brown.edu/faculty/Oded_Galor/pdf/Klien%20lecture.pdf

Galor, O. (2011). *Unified Growth Theory*. Princeton, New Jersey: Princeton University Press.

Galor, O. (2012). Unified Growth Theory and Comparative Economic Development. http://www.biu.ac.il/soc/ec/students/mini_courses/6_12/data/UGT-Luxembourg.pdf

Maddison, A. (2001). *The World Economy: A Millennial Perspective*. Paris: OECD.

Maddison, A. (2010). Historical Statistics of the World Economy: 1-2008 AD. http://www.ggdc.net/maddison/Historical Statistics/horizontal-file_02-2010.xls.

Nielsen, R. W. (2014). Changing the Paradigm. *Applied Mathematics*, *5*, 1950-1963. http://dx.doi.org/10.4236/am.2014.513188

Nielsen, R. W. (2015a). Mathematical Analysis of the Historical Economic Growth. http://arxiv.org/ftp/arxiv/papers/1509/1509.06612.pdf

Nielsen, R. W. (2015b). The Insecure Future of the World Economic Growth. *Journal of Economic and Social Thought*, *2*(4), 242-255.

Nielsen, R. W. (2015c). Early Warning Signs of the Economic Crisis in Greece: A Warning for Other Countries and Regions. http://arxiv.org/ftp/arxiv/papers/1511/1511.06992.pdf

Pereira, E. (2011). Developing Countries Will Lead Global Growth in 2011, Says World Bank. http://www.forbes.com/sites/evapereira/2011/01/12/developing-countries-will-lead-global-growth-in-2011-says-world-bank/

UNCTAD, (2013). Map of the LDCs. http://unctad.org/en/Pages/ALDC/Least%20Developed%20Countries/LDC-Map.aspx

UNDATA, (2015), Western Sahara. http://data.un.org/CountryProfile.aspx?crName=Western%20Sahara